\documentstyle[12pt,a4]{article}
\def\slash#1{#1\hskip -0.5em/}
\newcommand{\vdag}{v\hskip -0.5em/}
\newcommand{\Ddag}{D \hskip -0.6em / }
\newcommand{\DDdag}{{\cal D}\hskip -0.6em/}

\title{On the Renormalization of\\
	Heavy Quark Effective Field Theory}
\author{
        {\sc Wolfgang Kilian \quad
             Thomas Mannel }
        \vspace*{5mm} \\
        Institut f\"ur Kernphysik \\
        Technische Hochschule Darmstadt \\
        Schlossgartenstr. 9, D--64289 Darmstadt  \\
        Germany }
\begin{document}
\maketitle
\vbox to 0pt{\vss
  \vbox to \textheight{
    \begin{flushright}
      {\large \hfill IKDA 93/23 \\
      \hfill hep-ph/9307307}
    \end{flushright}    } }
\vfill
\begin{abstract}
\noindent
The construction of
heavy quark effective field theory (HqEFT) is extended to arbitrary order
in both expansion parameters $\alpha_s$ and $1/m_q$. Matching conditions
are discussed for the general case, and it is verified
that this approach correctly reproduces the infrared behaviour of full QCD.
Choosing a renormalization scheme in the full theory fixes the
renormalization scheme in the effective theory except for the scale
of the heavy quark field. Explicit formulae are given for the
effective Lagrangian, and one--loop matching renormalization
constants are computed for the operators of order $1/m$.
Finally, the multiparticle sector of HqEFT is considered.
\end{abstract}
\thispagestyle{empty}

\section{Introduction}
The Heavy Quark Limit (HQL) \cite{HQL} has become a very useful
tool for the description of systems involving one heavy quark. The main
progress consists in the exploitation of additional symmetries occuring
in the limit of infinite mass of heavy quarks. These additional
symmetries allow model independent predictions for systems with heavy
quarks, e.g., they yield model independent relations between form
factors of weak transition matrix elements and also model independent
absolute and relative normalizations of form factors at certain
kinematic points.

The predictions obtained in the HQL receive corrections which are
governed by two small parameters. The first type of corrections are
the QCD short distance corrections which are obtained in a perturbation
series in the parameter $\alpha_s (m)$, the strong coupling constant
at the scale of the mass $m$ of the heavy quark. The second type are the
recoil corrections governed by the small parameter $\Lambda_{\rm QCD}/m$.

A convenient tool to deal with these
corrections to the HQL is the so called Heavy Quark Effective Field
Theory (HqEFT) which was originally formulated in \cite{Gr90,Ge90,FG90}.
The construction of any effective theory consists of two steps \cite{Ge93}.
The first step is to identify the degrees of freedom which are
irrelevant at the scales under consideration and to remove them.
In the language of functional integrals
these degrees of freedom are integrated out in the functional integral.
Although generally
the action of the full theory is the integral of a local
Lagrangian, this first step will leave us with a nonlocal
action functional. However, this nonlocality is connected to the large
mass $M$ of the irrelevant degrees of freedom and an effective theory
may be constructed, if the nonlocal action functional can be expanded
into local terms where the expansion parameter is $1/M$.

This point of view has been elaborated in \cite{MRR92} for the
case of HqEFT by identifying the massive degees of freedom and by
explicitly integrating them out from the functional integral of QCD.
In this way the tree--level contributions in all orders of $1/m$
have been obtained. In other words, this approach yields the correct
HqEFT at the scale of the heavy quark mass $m$, i.e., the tree--level
matching of
full QCD to the effective theory. However, computing the coefficients
in the HqEFT Lagrangian to higher order in the perturbation expansion
necessarily involves loop calculations using the Feynman
rules of HqEFT. In the past these matching contributions usually
have been obtained by evaluating corresponding
diagrams in the full and effective theory separately, which is inconvenient
because of spurious infrared (IR) divergences
that cancel only in the final result.

The second problem with this
approach is connected to the multiparticle states of HqEFT. In the
HQL the numbers of heavy quarks and heavy antiquarks are separately
conserved and the derivation given in \cite{MRR92} only deals
with the one-particle sector of HqEFT. However, it has been noticed
that in the two-particle sector some unusual features of HqEFT appear
which are related to the fact that the naively calculated anomalous
dimensions pick up an imaginary part which leads to phases in the
Wilson coefficient functions. It has been shown later \cite{KMO93} that
one may in fact {\it define} real anomalous dimensions by redefining the
multiparticle states of HqEFT in an appropriate way.

The purpose of the present paper is to clarify these two points and
to supplement the discussion given in \cite{MRR92}. We shall
extend the work in \cite{MRR92} by stating the matching
and renormalization conditions for the effective theory beyond tree level
and thus
provide a method to construct the effective theory to arbitrary
order in the two expansion parameters. The ideas of \cite{Gr90}
are extended in order to show that this theory indeed reproduces
the results of full QCD in the given approximation.
In Sec.~2 we fix our notation by
showing at the level of Greens functions that the Lagrangian given
in \cite{MRR92} correctly
reproduces the QCD Greens functions at tree level. In Sec.~3 we
include also loop effects and discuss the matching
of HqEFT to full QCD. It is possible to summarize all corrections
in an effective Lagrangian which takes a simple form,
and thus to provide explicit formulae
for the matching coefficients to all orders. As an application,
in Sec.~4 we calculate the one--loop matching coefficients of
the operators in the Lagrangian up to order $1/m$.
Finally we address the problems of multiparticle
states in HqEFT and conclude.

\section{HqEFT at Tree Level}
We shortly review the derivation of the HqEFT at tree level as
it has been given in \cite{MRR92}. The heavy quark fields occuring
in the full QCD Lagrangian
\begin{equation}
  {\cal L} = \bar\psi(i\Ddag-m)\psi
  + \bar\eta\psi + \bar\psi\eta
\end{equation}
are rewritten using the projections on upper and lower components
with respect to the reference frame given by the velocity vector $v$
\begin{equation}
  P_v^\pm = \frac{1\pm \vdag}{2}
\end{equation}
as
\begin{eqnarray}
  P_v^+ \psi(x)
  &=& e^{-imv\cdot x}   h_v(x),
  \\
  P_v^-\psi(x)
  &=& e^{-imv\cdot x} H_v(x).
\end{eqnarray}
With the corresponding parameterization of the heavy quark sources
\begin{eqnarray}
  \rho_v(x) &=& P_v^+ e^{imv\cdot x}\eta(x), \\
  R_v(x)    &=& P_v^- e^{imv\cdot x}\eta(x)
\end{eqnarray}
and after integration over the $H_v$ fields
in the functional integral,
the tree--level Lagrangian becomes  a nonlocal expression
\begin{eqnarray}\label{L-tree-nl}
  {\cal L}^{(0)}_v &=&
  \bar h_v i(v\cdot D)h_v
  + (\bar h_viD_v^\perp + \bar R_v)
    \frac{1}{2m + i(v\cdot D)-i\epsilon}
    (iD_v^\perp h_v + R_v)  \nonumber\\
  && +\; \bar\rho_v h_v + \bar h_v\rho_v,
\end{eqnarray}
where the transverse derivative is given by
\begin{equation}
  D_v^\perp = \Ddag - \vdag(v\cdot D).
\end{equation}
The propagator of the heavy quark field can be read off as
\begin{equation}
  S_v^+(k) = \frac{iP_v^+}{v\cdot k + i\epsilon}
\end{equation}
and contains only forward propagation in time.

The nonlocal effective Lagrangian may be expanded in orders of $1/m$ as
\begin{eqnarray}\label{L-tree}
  {\cal L}_v^{(0)} &=&
  \bar h_v i(v\cdot D)h_v
  + (\bar h_vi\Ddag + \bar R_v)\frac{-i}{2m} P_v^-
    \sum_{n=0}^\infty \left(-\frac{iv\cdot D}{2m}\right)^n
    i(i\Ddag h_v + R_v)  \nonumber\\
  &&  +\; \bar\rho_v h_v + \bar h_v\rho_v,
\end{eqnarray}
where the backward propagation is lost if the series
is truncated at any finite order in $1/m$.
In these expressions we have retained the sources $R_v$ of the
lower component fields. They may be dropped if only Greens functions
sandwiched between $P_v^+$ projectors are to be calculated, which
is true if the external states are eigenstates of the lowest order
HqEFT Lagrangian. However,
if the external state contains a $P_v^-$ projection,
differentiations with respect to $R_v$ are necessary.

It may be verified that the Lagrangian (\ref{L-tree}) correctly
reproduces the tree level Greens functions of the full QCD Lagrangian
to arbitrary order in the $1/m$ expansion. To see this,
we rewrite the geometric series in (\ref{L-tree}) as
\begin{equation}\label{ges}
  \frac{-i}{2m}
  P_v^- \sum_{n=0}^\infty \left(-\frac{iv\cdot D}{2m}\right)^n P_v^-
  = S_v^- + S_v^-(i g\slash A) S_v^-
  + S_v^-(i g\slash A) S_v^- (i g\slash A) S_v^- + \ldots,
\end{equation}
where the expression $S_v$ which becomes in momentum space
\begin{equation}
  S_v^-(k) = -iP_v^-\frac{1}{2m}
	\sum_{n=0}^\infty\left(-\frac{v\cdot k}{2m}\right)^n
  = \frac{-iP_v^-}{2m+v\cdot k}
\end{equation}
may be interpreted as the propagator of the lower component field.
To any finite order in the $1/m$ expansion this is a local expression
and therefore it is part of the interaction terms in the effective
Lagrangian.

Using the Feynman rules as derived from (\ref{L-tree}), with the
upper component part of the Lagrangian in the form
\begin{equation}
  \bar h_v i(v\cdot D)h_v
  = \bar h_v P_v^+ i(v\cdot\partial)P_v^+ h_v
  + \bar h_v P_v^+(g\slash A) P_v^+ h_v,
\end{equation}
where we have made explicit the projection operators, we may
collect all terms that can occur in a Feynman diagram between
two insertions of the gluon field $g\slash A$. If both vertices
contain $P_v^+$ projectors, they sum up to
\begin{eqnarray}
  G_v^{++} &=&
  S_v^+
  + S_v^+ \left(i\slash k S_v^- i\slash k\right) S_v^+
  + S_v^+ \left(i\slash k S_v^- i\slash k\right) S_v^+
	\left(i\slash k S_v^- i\slash k\right) S_v^+  + \ldots \nonumber\\
  &=& iP_v^+ \frac{1 + v\cdot k/2m}{v\cdot k + k^2/2m}.
\end{eqnarray}
Similarly, if one insertion contains a $P_v^-$ projection,
we have
\begin{eqnarray}
  G_v^{+-} &=& G_v^{++}i\slash k S_v^- \nonumber\\
  &=& iP_v^+ \frac{\slash k/2m}{v\cdot k + k^2/2m} P_v^-,
\end{eqnarray}
whereas between two $P_v^-$ projectors there can be an additional
term from the geometric series (\ref{ges})
\begin{eqnarray}
  G_v^{--} &=& S_v^- i\slash k G_v^{+-} + S_v^- \nonumber\\
  &=& -i P_v^-\frac{v\cdot k/2m}{v\cdot k + k^2/2m}.
\end{eqnarray}
In the sum of the four possible contributions the projection operators
disappear
\begin{equation}
  G_v^{++} + G_v^{+-} + G_v^{-+} + G_v^{--}
  = i\frac{m\vdag + \slash k + m}{(mv + k)^2-m^2}
\end{equation}
and the propagator of the full theory is recovered. The same is
true if one or both of the gluon field insertions are replaced
by external sources. Note that in the latter case the term quadratic
in $R_v$ becomes important.

The $1/m$ expansion of a tree--level Greens function is unique. We have
seen that the full propagator, and thus any Greens function, is correctly
reproduced by the effective theory. Truncating the series in the
effective Lagrangian at a given order $1/m^k$, we get an approximation
to the full Greens function which differs by terms of order $1/m^{k+1}$:
\begin{equation}\label{tree-matching}
  G_{hl}^{(0)}(\lambda,m)
	- \tilde G_{hl}^{(0)}(\lambda, 1/m)
  = O\left(g^{h+l-2}
    \lambda^{\delta}(\lambda/m)^{k+1}\right),
\end{equation}
where $\delta$ is the mass dimension of the Greens function, and
the momenta of the $h$ external heavy and $l$ external massless
lines are parameterized by
\begin{eqnarray}
  P_i &=& mv + \lambda x_i, \nonumber\\
  p_i &=& \lambda y_i,
\end{eqnarray}
so that the parameter $\lambda$ measures the deviation of the
external momenta from mass shell.

Although these facts might seem trivial for the tree--level case,
we note that some parts of the full theory propagator shrink to
a point and become
part of the interaction vertices in the effective theory. Thus,
the one--particle irreducible (1PI) Greens functions of the full theory
do not correspond to the 1PI functions in the effective theory.
This fact has to be accounted for in the matching of higher dimensional
operators at order $\alpha_s$ or higher.

\section{Matching and Renormalization Conditions}
If we want a statement like (\ref{tree-matching}) to hold at higher
orders in the loop expansion, we have to include in the
effective Lagrangian all possible
local operators up to a given mass dimension and heavy--quark number
that cannot be excluded by symmetry arguments.
Furthermore, the operators that are already
present in the tree--level Lagrangian will be modified by multiplicative
renormalizations which in general are ultraviolet (UV)
divergent. However, no IR infinities can occur if the calculations are
organized appropriately, so one can dispense of all IR regulators in
the matching calculations.

In comparing the one loop Greens functions of the full and
effective theories we first consider the Greens functions
which contain no external heavy quark. Diagrams without
any internal heavy quark line are identical in both
theories. On the other hand, diagrams with heavy quark loops have
no counterpart in the effective theory. This is true for any
diagram even if two heavy quarks with different velocities are
present: The heavy quark propagator in coordinate space
\begin{equation}
  S_v^+(x) = \theta(x_0)\,\delta^3(\vec x - \vec v x_0/v_0)
\end{equation}
describes a particle moving along a classical straight world line
in the forward light cone. If two heavy propagators are connected
at one point (the origin) in coordinate space, the loop cannot be
closed at any other point, so the loop integral vanishes.
To be precise,
because at the origin two distributions are multiplied, it is
equal to some undefined constant which can be got rid of by
a simple renormalization. In contrast to light particles, a
heavy particle loop cannot introduce any nonlocal interaction.

Thus, one can calculate the coefficients of operators
without heavy quarks such as
\begin{equation}\label{det}
  \frac{1}{(2m)^2}(D^\mu G_{\mu\rho}) (D_\nu G^{\nu\rho}), \quad
  \frac{1}{(2m)^2}G^\mu_{\,\nu} G^\nu_{\,\rho} G^\rho_{\,\mu}.
\end{equation}
by computing the corresponding
loop diagrams in the full theory. This is just the result
we would have obtained if we had integrated out the heavy quark
completely. At one loop the result is given by the determinant
of the heavy fermions in the functional integral \cite{BO86}.

For Greens functions
with external heavy quarks we need additional counterterms.
These are to be chosen in such a way that the condition
analogous to (\ref{tree-matching}) for $n$ loop
Greens functions with $h$ external heavy--quark and $l$
external gluon lines\footnote{
	Throughout this paper we ignore the ghost fields, and
	do not write out the pure gluon and gauge fixing terms
	in the Lagrangian. The ghost fields can be treated in the
	matching in the same way as the gluon fields. To order
	$1/m^0$ the question of BRS invariance has been addressed
	in \cite{BG93}. A complete discussion
	to all orders is beyond the scope of this paper.}
\begin{equation}\label{loop-matching}
  \Delta_{hl}^{(n)}(\lambda,m)
  = G_{hl}^{(n)}(\lambda,m)
	- \tilde G_{hl}^{(n)}(\lambda, 1/m)
  = O\left(g^{h+l-2}\alpha_s^n
    \lambda^{\delta}(\lambda/m)^{k+1}\ln^n(\lambda/m)\right),
\end{equation}
can be satisfied for $h\leq h_0$,
if we know that it is already true in lower order of the loop
expansion. For the lowest order ($1/m^0$) Greens functions with $h_0=2$
this has
been shown in \cite{Gr90}. Extending the ideas developed there we
shall show in the following paragraphs how the matching can
be accomplished to arbitrary order in the $1/m$ expansion.

For $\delta+k\geq 0$, (\ref{loop-matching}) is equivalent to the requirement
that the first $\delta+k$ derivatives of
$\Delta_{hl}^{(n)}(\lambda,m)$ with respect to external
momenta vanish for $\lambda\to 0$. If the quantity
$\hat\Delta_{hl}^{(n)}(\lambda,m)$, which is defined in the same way as
$\Delta_{hl}^{(n)}(\lambda,m)$ but includes counterterms only
up to $n-1$ loop order, admits an expansion of the form
\begin{eqnarray}\label{loop-expansion}
  \hat\Delta_{hl}^{(n)}(\lambda,m)
  &=& C_0 m^\delta + C_1 \lambda m^{\delta-1} + \ldots
    + C_{\delta+k}\frac{\lambda^{\delta+k}}{(2m)^k}
  \nonumber\\
  &&  + \;O\left(g^{h+l-2}\alpha_s^n
    \lambda^{\delta}(\lambda/m)^{k+1}\ln^n(\lambda/m)\right),
\end{eqnarray}
this can be accomplished by introducing
local counterterms of order $\alpha_s^n$ and up to order $1/m^k$
into the effective Lagrangian.
Since $\omega=\delta+k$ is the maximal
UV degree of divergence of the effective theory diagrams, the
coefficients $C_0$ to $C_{\delta+k}$ which translate into
coefficients of counterterms are in general UV divergent.
To make (\ref{loop-matching}) hold,
we have to include counterterms exactly for those diagrams
with $\omega\geq 0$, so the set of matching conditions (\ref{loop-matching})
uniquely defines a renormalization scheme in the effective theory.

If (\ref{loop-expansion}) did not hold, we would encounter
IR divergences in evaluating $\hat\Delta_{hl}^{(n)}$ or its
derivatives for vanishing external momenta. These could come
from regions where one or more loop momenta become small.
We consider a particular diagram of the full theory together
with its counterparts in the effective theory up to
order $1/m^k$, including all counterterm insertions up to
$n-1$ loop order. Let us investigate the behaviour of the integrand
in the region where the momenta of some subset of $s$ massless propagators
become small of order $\lambda$. Since the $s$ light propagators are the same
in both theories, we can factor them out in the difference. The remaining
expression corresponds to a Greens function --- not necessarily
connected --- in
lower order of the loop expansion. It is represented by the same set of
diagrams, where the $s$ light lines have been cut, their momenta being
regarded as external. If the remaining $n'$
loop momenta are integrated over, according
to the induction hypothesis this Greens function satisfies
\begin{equation}
  \Delta_{hl'}^{(n')}(\lambda,m)
  = O\left(g^{h+l-2}\alpha_s^{n'}
	\lambda^{\delta'}(\lambda/m)^{k+1}\ln^{n'}(\lambda/m)\right),
\end{equation}
where $\delta' = \delta+2s-4(n-n')$. The total IR degree
of divergence in this integration region is therefore
\begin{equation}
  \rho = - [(\delta' + k + 1) -2s + 4(n-n')] = -\delta-k-1,
\end{equation}
and thus no IR singularities show up
in the first $\delta+k$ derivatives of $\hat\Delta_{hl}^{(n)}$.

Any diagram with $\delta+k<0$ automatically satisfies (\ref{loop-matching})
without additional counterterms, since by the same reasoning
the matching conditions
for its subdiagrams exclude unexpected positive powers of
$m$ coming from UV divergent subintegrations. Simple power counting
then directly gives (\ref{loop-matching}) for these diagrams.

We did not consider additional IR divergences from integration
regions where some momenta of heavy quark lines become small.
In fact there are none. If we cut a heavy quark line, the remaining
subdiagram has a larger number of external heavy quark lines and
thus it is part of a different sector of the theory. Violations of
the IR power counting could occur if this subdiagram is
UV divergent, which can happen first at order $1/m^2$.
We are tempted to renormalize it by imposing matching conditions
also for Greens functions with more than $h$ external heavy lines, but
as explained in Sec.~5, their counterterms can contain divergent phases
which should not appear in the Lagrangian. Fortunately,
these terms give vanishing contributions to the final result, because
in evaluating them, we have to close the heavy quark line again
and get a heavy quark loop which is zero by definition.
We conclude that the multiparticle sectors are completely irrelevant
for the renormalization and matching of the one--particle sector,
and all possible IR singularities have been accounted for in the
previous paragraphs.

In practice, the counterterms needed in the effective Lagrangian
are given by the $1/m$ expansion up to a certain order $1/m^k$
of the 1PI diagrams in the
full theory, with all corresponding diagrams of the effective
theory up to the same order subtracted. Since the effective theory
does not contain any mass scale, the subtraction terms
are defined uniquely by the requirement that the first $\delta+k$
derivatives of the subtracted diagram are IR finite, where $\delta$
is the mass dimension of the diagram, so that the
effective theory is not needed explicitely in the matching calculation.

In evaluating the counterterms introduced by a particular 1PI diagram
into the effective theory, we first subtract all counterterm
diagrams necessary
to make this diagram UV finite in the full theory. These may be taken in the
$\overline{\rm MS}$ scheme as usual, with one exception: In
order to have a renormalization group invariant mass as expansion
parameter, we must also include a counterterm for the finite
part of the quark mass renormalization. In this way the self--energy
diagram contains no constant term, and the $1/m$ expansion is
done in terms of the pole mass which has to be redefined in each
order of the perturbation expansion. Furthermore, the counterterms
in the effective Lagrangian all have nonpositive mass dimension\footnote{
	The possibility of expanding in terms of a different mass,
	which introduces a residual mass term into the effective
	Lagrangian, has been considered in \cite{FLN92}.}.

We may put the results of the preceding paragraphs together in a formal
language,
and give an explicit expression for the effective theory counterterm
of a particular diagram $\Gamma$ with mass dimension $\delta(\Gamma)$.
For a one--loop diagram, it
is given by the integral of the expression
\begin{equation}\label{bog}
  C_\Gamma = \tilde{\cal T}_{\delta(\Gamma)+k}I_\Gamma
	= (1 - {\cal J}_\rho){\cal T}_{\delta(\Gamma)+k}I_\Gamma,
\end{equation}
where $I_\Gamma$ is the integrand of the full theory diagram. The
operator ${\cal T}_{\delta+k}$ gives the Taylor expansion of the integrand
in terms of {external} momenta up to the required order $\delta+k$.
The operator ${\cal J}_\rho$ gives the first $\rho+1$ terms of a Laurent
expansion
in terms of {\em internal} momenta, where $\rho$ is the IR degree of
divergence of the term on which ${\cal J}_\rho$ acts. This takes into
account the diagrams of the effective theory. All expansions are done
about zero momentum $k=0$, where the heavy quark
momenta are parameterized by $p=mv + k$.

Because of the additional
subtractions, the subtracted integrand $C_\Gamma$ is no longer
UV finite, but has an UV degree of divergence $\omega=\delta+k$.
Since it is a polynomial in external momenta of order $\delta+k$
which is designed to cancel the leading terms in the difference of the full and
effective theory diagrams, it also exactly cancels the UV divergences
in this difference.

Evaluating the counterterms for $n$--loop graphs, we have to take
into account all lower order counterterms in the effective theory.
We observe that a recursive procedure emerges
closely analogous to the well--known renormalization of UV divergences.
Thus we may give the effective theory counterterm of
any $n$--loop 1PI diagram
in closed form as an extension of Zimmermann's forest formula \cite{Zi73}
\begin{equation}\label{zim}
  C_\Gamma =
	\tilde{\cal T}_{\delta(\Gamma)+k}
	\sum_{F:\;\Gamma\not\in F}\prod_{\gamma\in F}
	\left(-\tilde{\cal T}_{\delta(\gamma)+k}\right)I_\Gamma,
\end{equation}
where the modified Taylor operator $\tilde{\cal T}$
is defined as in (\ref{bog}),
and the sum runs over all forests $F$ of subdiagrams $\gamma$ not
containing $\Gamma$ itself. The ``renormalized'' integrand
\begin{equation}
  R_\Gamma =
	\sum_{{\rm all}\; F}\prod_{\gamma\in F}
	\left(-\tilde{\cal T}_{\delta(\gamma)+k}\right)I_\Gamma
\end{equation}
is the full theory diagram with all effective
theory diagrams subtracted, i.e., a term contributing to the
difference $\Delta_{hl}^{(n)}$ which obeys the matching condition
(\ref{loop-matching}). Of course, we can apply these formulae also
to the $h=0$ case, where the IR subtractions first appear
in two--loop order.

This treatment of the matching conditions is not just formal.
In fact, it is convenient because all integrations are manifestly
IR finite, and the integrals can be evaluated for vanishing external
momenta. In dimensional regularization, the $1/\epsilon^n$ poles
are due only to UV singularities.

In the effective theory we have the freedom left to rescale the
heavy quark field by a finite amount $z$. The
matching condition (\ref{loop-matching}) is then modified to
\begin{equation}\label{ms-loop-matching}
  G_{hl}^{(n)}(\lambda,m)
	- z^{h/2}\tilde G_{hl}^{(n)}(\lambda, 1/m)
  = O\left(g^{h+l-2}\alpha_s^n
    \lambda^{\delta}(\lambda/m)^{k+1}\ln^n(\lambda/m)\right).
\end{equation}
If the renormalization scheme of the full theory is fixed, choosing
a value for $z$ and requiring that (\ref{ms-loop-matching}) holds
uniquely defines a renormalization scheme for the
effective theory. Using the $\overline{\rm MS}$ scheme
in the lowest order effective
theory amounts to specifying a wave function renormalization constant $z$
that is given by
the finite part of the difference of the heavy quark self energy
diagrams in the full and effective theory.
In higher order of the $1/m$ expansion we have to insert the
counterterms as given by (\ref{zim}) explicitely, with the self--energy
contribution of the external lines factored out.

We may collect all integrands $C_\Gamma$ (\ref{zim}) that contribute
to the counterterm of some 1PI Greens function with $h$ heavy--quark
and $l$ gluon legs. After integrating over internal momenta,
this becomes a function $C_{hl}$ which is a polynomial of order
$\delta+k$ in external momenta. Fourier transforming back into
coordinate space and inserting gluon fields (we do not write out
colour indices), we define the quantity
\begin{equation}\label{full-matching}
  {\cal C}_h = \sum_l \frac{1}{l!} C_{hl}^{\mu_1\cdots\mu_l}(i\partial)\,
	A_{\mu_1}\cdots A_{\mu_l}
\end{equation}
which looks like a generating functional of 1PI vertices, but does not
contain the heavy quark fields yet. In contrast to the effective
action of the full theory, it can be expanded as a series
of powers of $\partial/m$.
By definition, we also include the tree--level contribution into
${\cal C}_2$ which is just the ordinary vertex $g\slash A$.

In the following paragraphs, we consider only the case $h\leq 2$,
the one--particle case. With the caveat discussed at the end of Sec.~5, the
generalization to multiparticle Greens functions is straightforward.

The effective vertices which are summarized in ${\cal C}_2$ can
be linked together by propagator projections in the same way
as the vertices $g\slash A$ at tree--level. Repeating the
reasoning of Sec.~2, we find that we can incorporate all loop
effects in the effective Lagrangian by defining a generalized
covariant derivative
\begin{equation}\label{D-gen}
  i\DDdag = i\slash\partial + {\cal C}_2 = i\Ddag + O(\alpha_s)
\end{equation}
with projections
\begin{eqnarray}\label{D-gen-p}
  i{\cal D}_v^+ &=& P_v^+i\DDdag P_v^+ = (iv\cdot D)P_v^+ +
O(\alpha_s),\nonumber\\
  i{\cal D}_v^\perp &=& P_v^+i\DDdag P_v^- + P_v^-i\DDdag P_v^+
	= iD_v^\perp + O(\alpha_s)\nonumber,\\
  i{\cal D}_v^- &=& P_v^-i\DDdag P_v^- = -(iv\cdot D)P_v^- + O(\alpha_s).
\end{eqnarray}
The effective Lagrangian which generalizes (\ref{L-tree-nl}) then reads
\begin{eqnarray}\label{L-loop}
  {\cal L}_v &=&
  \bar h_v i{\cal D}_v^+ h_v
  + (\bar h_vi{\cal D}_v^\perp + \bar R_v)
    \frac{1}{2m-i{\cal D}_v^- }
    (i{\cal D}_v^\perp h_v + R_v) +{\cal C}_0 \nonumber\\
  && +\;\bar\rho_v h_v + \bar h_v\rho_v,
\end{eqnarray}
where ${\cal C}_0$ summarizes the counterterms without heavy quarks.
This effective Lagrangian is valid to arbitrary order in the $1/m$
and loop expansions.

It is straightforward to apply this procedure to operator insertions.
For instance, a heavy--light current $K=\bar q\Gamma\psi$, where $q$
is some light quark field, is matched onto
\begin{equation}\label{O-matching}
  K_v = \bar q {\cal K}\left[
	\frac{1}{2m-i{\cal D}_v^- }
    (i{\cal D}_v^\perp h_v + R_v) + h_v \right],
\end{equation}
where ${\cal K}$ stands for the sum of all 1PI diagrams of the full
theory that involve one insertion of $K$, with an arbitrary number
of gluons, where the IR subtractions as in (\ref{zim}) have been
carried out.

The form of ${\cal L}_v$ and $K_v$ is restricted if BRS
invariance holds in the matching. One would expect the
generalized covariant derivative $\DDdag$ to depend on the
gluon field only through the ordinary covariant derivative
$i\Ddag = i\slash\partial + g\slash A$, at least in the
background field gauge \cite{Ab81}.
This has been shown for the
lowest order ($1/m^0$) theory in \cite{BG93}.

Furthermore, there
are restrictions following from reparameterization invariance \cite{LM92}.
They emerge from the fact that the Greens functions of the
full theory do not depend on the heavy--quark velocity $v$ and the
residual momentum $k$ separately, but only in the combination
$p=mv+k$. In the effective theory, the Lagrangian therefore depends
on the velocity and the covariant derivative only in the combination
${\cal V} = v + iD/m$. This introduces relations among operator
coefficients in different orders of the $1/m$ expansion.

\section{Coefficients to order $1/m$ and $\alpha_s$}

As an example, in this section we shall discuss the one--loop matching
up to order $1/m$. By combining denominators and algebraically reducing
vector and tensor integrals, all quantities that occur in the one--loop
matching calculations to arbitrary order in the $1/m$ expansion
can be expressed in terms of IR--finite integrals in the form
\begin{equation}
  \Xi_{abc}(m) = \int dk \frac{1}{(k^2)^a (v\cdot k)^b (v\cdot k + k^2/2m)^c},
\end{equation}
of which only a few have to be evaluated because of the relations
\begin{equation}\label{xi-abl}
  \frac{d}{dm}\Xi_{abc}(m) = \frac{c}{2m^2} \Xi_{a-1,b,c+1}(m)
\end{equation}
and
\begin{equation}\label{xi-dif}
  \Xi_{abc}(m) = 2m\left(\Xi_{a+1,b,c-1}(m)-\Xi_{a+1,b-1,c}(m)\right).
\end{equation}

To order $1/m$ there are three independent operators
involving two heavy quarks possible which are usually written in
the form
\begin{equation}\label{1/m}
  \frac{1}{2m}\bar h_v (iD)^2 h_v, \quad
  \frac{1}{2m}\bar h_v (iv\cdot D)^2 h_v, \quad
  \frac{g}{4m}\bar h_v \sigma^{\mu\nu}G_{\mu\nu} h_v.
\end{equation}
The second operator vanishes when sandwiched between physical states
due to the heavy--quark equations of motion.

Because of BRS invariance one only has to
calculate the two one--loop vertex diagrams of the full theory,
with the appropriate IR subtractions, to obtain the
matching coefficients up to order $\alpha_s$ and $1/m$.
Applying the formulae
(\ref{bog},\ref{full-matching},\ref{L-loop}), and expanding
the result up to order
$1/m$, we obtain the one--loop effective Lagrangian at the
matching scale $\mu=m$ in the background field gauge with $\xi=1$
(Feynman gauge)
\begin{equation}\label{L-1/m}
  {\cal L}_v^{(1)} = Z_\psi^{-1}Z_h \bar h_v\left[
	iv\cdot D + Z_1\frac{(iD)^2}{2m} - Z_2\frac{(iv\cdot D)^2}{2m}
	+ Z_3 \frac{g}{4m}\sigma_{\mu\nu}G^{\mu\nu}\right]
	h_v,
\end{equation}
where the matching renormalization constants are given by
[$1/\hat\epsilon = 1/\epsilon + \ln(4\pi)/2- \gamma_E/2 $,
where the space--time dimension is $4-\epsilon$]
\begin{eqnarray}\label{L-match}
  Z_h &=& 1 + \frac{\alpha_s}\pi\left(\frac{1}{\hat\epsilon}+1\right)C_F,
\nonumber\\
  Z_1 &=& 1, \nonumber\\
  Z_2 &=& 1 + \frac{\alpha_s}\pi\left(\frac{3}{\hat\epsilon}+\frac12\right)C_F,
\nonumber\\
  Z_3 &=& 1 + \frac{\alpha_s}\pi\left[
	\frac12 C_F + \left(\frac{1}{2\hat\epsilon}+\frac12\right)C_A\right],
\end{eqnarray}
and the wave--function renormalization of the full theory is in the
$\overline{\rm MS}$ scheme
\begin{equation}
  Z_\psi = 1 + \frac{\alpha_s}\pi\left(-\frac{1}{2\hat\epsilon}\right)C_F .
\end{equation}
The values of $Z_h$, $Z_1$, and $Z_2$ are also obtained by computing
the one--loop self--energy to the required order. In addition, this
gives the mass renormalization
\begin{equation}
  \delta m = \frac{\alpha_s}\pi C_F\left(-\frac{3}{2\hat\epsilon}-1\right)m_0
\end{equation}
so that the $1/m$--expansion is done in terms of the pole mass
\begin{equation}
  m = m_0 - \delta m.
\end{equation}

The fact that $Z_1=1$ is a consequence of reparameterization invariance
\cite{LM92}. BRS invariance ensures that no gauge--variant operators
appear, and that only the unphysical constants $Z_\psi$, $Z_h$, and
$Z_2$ depend on the gauge parameter. Our results for the one--loop
matching renormalization constants
agree with those given in \cite{EH90}. The calculational
method of \cite{EH90}, where dimensional regularization has been used
for both UV and IR singularities, is thus justified by employing a scheme that
is
manifestly free of IR divergences.

In order to obtain the effective Lagrangian valid at scales $\mu<m$,
one may sum up the leading mass logarithms by
evaluating the UV--divergent parts of the loop diagrams
in the effective theory. The leading log calculation for the
Lagrangian (\ref{L-1/m}) has been performed in \cite{FGL91}. In fact,
the one--loop anomalous dimensions are determined by the
$1/\epsilon$--poles of the renormalization constants (\ref{L-match}).
The finite terms in (\ref{L-match}) become relevant when
subleading logarithms are summed.

\section{Hadronic States}
In a typical application of HqEFT one considers the matrix element
of some operator ${\cal O}$, describing e.g.\ the weak decay of a
heavy quark, between hadronic states $|{\rm in}\rangle$ and
$|{\rm out}\rangle$ involving heavy quarks. As in ordinary perturbation
theory, the full QCD eigenstates are evolved from the corresponding
lowest order HqEFT eigenstates $|{\rm in(out) H}\rangle$ by adiabatically
switching on the $1/m$ terms in the effective Lagrangian:
\begin{equation}\label{GML}
  \langle{\rm out}|{\cal O}|{\rm in}\rangle
  = \langle{\rm out\; H}|\,{\rm T}\left[
	\exp(-iS'_v)\,\tilde{\cal O}\right]|{\rm in\; H}\rangle,
\end{equation}
where $S'_v = \int dx\,{\cal L}'_v(x)$ is given by the effective
Lagrangian (\ref{L-loop}) excluding terms of order $1/m^0$, and $\tilde{\cal
O}$
is the result of a matching calculation such as (\ref{O-matching}).

Usually, (\ref{GML}) cannot be evaluated completely in terms of Feynman
diagrams because perturbation theory breaks down at low scales. One
rather uses perturbation theory to sum up the leading (and subleading)
logarithms $\log(m/\mu)$, where $\mu$ is some low hadronic scale,
and expresses the result in terms of some operator ${\cal O}(\mu)$
which has the same matrix element between scaled--down states
$|{\rm in(out) H},\mu\rangle$. These are defined in such a way that
the matrix element is $\mu$ independent:
\begin{equation}\label{GML2}
  \langle{\rm out\; H}|\,{\rm T}\left[
	\exp(-iS'_v)\,\tilde{\cal O}\right]|{\rm in\; H}\rangle
  = \langle{\rm out\; H},\mu|{\cal O}(\mu)|{\rm in\; H,\mu}\rangle.
\end{equation}
Obviously, ${\cal O}(\mu)$ contains nonlocal operators
(time--ordered products) in its $1/m$ expansion. In the solution
of the renormalization group equations they mix with local operators.

If we want to include into this treatment also states involving
more than one heavy quark at a time, we have to extend the
heavy--quark Lagrangian accordingly. We can apply HqEFT
to asymptotic states
where the heavy quarks end up in different hadrons, so each
one can be assigned a separate velocity. The Lagrangian
then contains a sum over all heavy quark and antiquark
flavours and velocities that occur in the problem
\begin{eqnarray}\label{L-multi}
  {\cal L} &=& \sum_{f,v}\left\{
  \bar h^f_v i{\cal D}_v^+h^f_v
  + (\bar h^f_vi{\cal D}_v^\perp + \bar R^f_v)
  \frac{1}{2m-i{\cal D}_v^-}
    (i{\cal D}_v^\perp h^f_v + R^f_v)\right.  \nonumber\\
  && \quad\quad\left.\vphantom{\frac{1}{{\cal D}_v^\perp}}
  +\; \bar\rho^f_v h^f_v   + \bar h^f_v\rho^f_v\right\}
  + \sum_f {\cal C}_0^f
  + \mbox{(counterterms with $h>2$)},
\end{eqnarray}
where the projections of the generalized covariant derivative
$\DDdag$ (\ref{D-gen}) have been defined in (\ref{D-gen-p}).
The counterterms without heavy quarks ${\cal C}_0$
are inserted only once for each quark flavour.
The bilinear matching corrections that are incorporated
in $\DDdag$ are needed
once for each flavor {\em and} velocity. In addition, one
expects counterterms involving more heavy quarks at higher
order in the $1/m$ expansion if there is more than one heavy
quark present in the initial or final states. These
counterterms can link different velocity and flavour sectors.
They are needed only if they involve no more heavy
quarks than the process under consideration:
Any diagram which contains an operator with more heavy
fields than there are available as external lines necessarily
contains a heavy quark loop and thus vanishes.

If more than one heavy quark is present at the same time,
some Feynman amplitudes develop UV divergent phases
analogous to the well--known Coulomb phases of QED \cite{GK91}.
They can be absorbed into the definition of
the multiparticle states itself \cite{KMO93}. In particular,
the phase of the two--particle final state is related
to the static interquark potential \cite{KMO93}.
If two velocities of different heavy quarks become
equal, these phases become infinite, so that the whole
expression is ill--defined. However, one could go back
into coordinate space and consider the case of two separated
heavy quarks which act as static coulour sources. This notion
was taken in the first perturbative evaluation of the interquark
potential \cite{Fis77}.

\section{Conclusions}
In the present paper we have extended the construction of HqEFT to arbitrary
order in both expansion parameters $1/m$ and $\alpha_s$. Although
our arguments as given in Sec.~2 are merely heuristic, we believe
that they suffice to show that the effective theory
reproduces the behaviour of full QCD near mass shell to any
required accuracy. The formulae (\ref{zim}) and (\ref{L-loop})
provide a method to obtain the terms in the effective Lagrangian
directly from the full theory. The fact that no IR regulators
need to be introduced simplifies practical calculations. As
an example, we have rederived the coefficients of the terms
in the effective Lagrangian up to order $1/m$ and $\alpha_s$.

For the construction of the complete effective Lagrangian (\ref{L-loop})
one needs in principle the knowledge of all diagrams of the full theory.
One might argue that for this reason the effective theory
is without physical content. However, its main power lies
in the possibility of summing up large logarithms of the
heavy quark mass, which are difficult to extract from the
full theory expressions, and thus reorganizing the perturbation
series. The method for calculating anomalous dimensions
and solving the renormalization group equations within the
effective theory  is well known. The Lagrangian (\ref{L-loop})
provides in closed form the initial conditions valid at the matching scale
$\mu=m$. It may also serve as a starting point for theoretical
considerations, e.g., the proof of BRS invariance to all orders.

In the last section we have considered the extension of HqEFT to
sectors with more than one heavy quark. It should have become
clear that these sectors do not introduce difficulties if
the heavy quarks remain well separated in the asymptotic states,
because
the UV--divergent phases that appear in these sectors can then be
absorbed in the definition of the multiparticle states.

All arguments in the present paper have been in the context
of perturbation theory. We did not adress the question whether
a nonperturbative approach to the $1/m$ expansion such as
lattice gauge theory introduces any new problems. However, the
perturbative matching of the continuum theory to the lattice
regulated version should proceed mainly along the same lines
as we have discussed.

\end{document}